\documentclass{sig-alternate-05-2015}
\setcopyright{acmcopyright}
\usepackage{acronym}
\usepackage{booktabs}
\usepackage{cite}
\usepackage{graphicx}
\usepackage{amsmath}
\usepackage{algorithm}
\usepackage[noend]{algpseudocode}
\usepackage[tight,footnotesize]{subfigure}
\usepackage{url}
\usepackage{etoolbox}
\usepackage{color}
\makeatletter
\patchcmd{\maketitle}{\@copyrightspace}{}{}{}
\makeatother

\newcommand{\circlenum}[1]{\raisebox{.5pt}{\textcircled{\raisebox{-.9pt} {\sf #1}}}}

\begin{document}

\title{Parallel Sort-Based Matching for Data Distribution Management on Shared-Memory Multiprocessors\footnotemark}

\numberofauthors{2}
\author{
\alignauthor
Moreno Marzolla\\
\affaddr{Dept. of Computer Science and Engineering}\\
\affaddr{University of Bologna, Italy}\\
\email{moreno.marzolla@unibo.it}
\alignauthor
Gabriele D'Angelo\\
\affaddr{Dept. of Computer Science and Engineering}\\
\affaddr{University of Bologna, Italy}\\
\email{g.dangelo@unibo.it}
}

\maketitle

\footnotetext{The publisher version of this paper is available at \url{https://doi.org/10.1109/DISTRA.2017.8167660}.
\textbf{{\color{red}Please cite this paper as: ``Moreno Marzolla, Gabriele D'Angelo. Parallel Sort-Based Matching for Data Distribution Management on Shared-Memory Multiprocessors. Proceedings of the IEEE/ACM International Symposium on Distributed Simulation and Real Time Applications (DS-RT 2017)''.}
{\color{blue}Best Paper Award @DS-RT 2017.}}}

\begin{abstract}
In this paper we consider the problem of identifying intersections
between two sets of~$d$-dimensional axis-parallel rectangles. This is
a common problem that arises in many agent-based simulation studies,
and is of central importance in the context of~High Level Architecture
(HLA), where it is at the core of the Data Distribution Management
(DDM) service. Several realizations of the DDM service have been
proposed; however, many of them are either inefficient or inherently
sequential. These are serious limitations since multicore processors
are now ubiquitous, and DDM algorithms -- being CPU-intensive -- could
benefit from additional computing power. We propose a parallel version
of the Sort-Based Matching algorithm for shared-memory
multiprocessors. Sort-Based Matching is one of the most efficient
serial algorithms for the~DDM problem, but is quite difficult to
parallelize due to data dependencies. We describe the algorithm and
compute its asymptotic running time; we complete the analysis by
assessing its performance and scalability through extensive
experiments on two commodity multicore systems based on a dual socket
Intel Xeon processor, and a single socket Intel Core i7 processor.
\end{abstract}

\ccsdesc[500]{Computing methodologies~Massively parallel and high-performance simulations}
\ccsdesc[300]{Computing methodologies~Shared memory algorithms}

\printccsdesc

\keywords{Data Distribution Management (DDM), Parallel And Distributed Simulation (PADS), High Level Architecture (HLA), Parallel Algorithms}

\section{Introduction}\label{sec:introduction}

Agent-based simulations involve a possibly large number of agents that
interact in a virtual environment. Generally, the environment may
represent a two- or three-dimensional space. For example, in a
large-scale road traffic simulation, agents may represent vehicles
moving in a two-dimensional, ``flat'' road network (the third
dimensions can be ignored since vehicles are concerned about obstacles
on their plane of movement only). Molecular models or air traffic
simulations, on the other hand, involve agents moving in a
three-dimensional world.

Agents must be made aware of events happening in their area of
interest, so that they can promptly react if necessary. For example,
in the road traffic scenario above, each car should be made aware of
the behavior of neighboring vehicles only, since distant vehicles can
not produce immediate observable effects. For simplicity, an agent's
area of interest is often represented as a $d$-dimensional rectangle
(\emph{region}), centered at the agent coordinates, with the sides
parallel to the axes of a $d$-dimensional space (usually, $d=2$ or
$d=3$). A simulation event that is generated by an agent~$A$ should
then be forwarded to all agents whose area of interest intersect that
of~$A$.

Managing areas of interest in agent-based simulations is so common
that the~\ac{HLA} specification~\cite{HLA} defines \ac{DDM} services
to handle the problem. Specifically, \ac{DDM} services are responsible
for sending events generated on \emph{update} regions to a set of
\emph{subscription} regions.

Identifying all pairs of intersecting rectangles is a well-known
computational geometry problem with applications in such diverse areas
as~VLSI design and geographic information systems. Spatial data
structures that can solve the region intersection problem have been
developed: examples include the $k$-$d$ tree~\cite{Rosenberg1985} and
R-tree~\cite{Guttman1984}. However, it turns out that \ac{DDM}
implementations tend to rely on less efficient but simpler
solutions. The reason is that spatial data structures can be difficult
to implement and their manipulation incurs a significant overhead
which is not evident from their asymptotic complexities.

The increasingly large size of agent-based simulations is posing a
challenge to the existing implementations of the \ac{DDM} service. As
the number of regions increases, so does the execution time of the
intersection-finding algorithms. A possible solution comes from the
computer architectures domain. The current trend in microprocessor
design is to put more execution units (cores) in the same processor;
the result is that multi-core processors are now ubiquitous, so it
makes sense to try to exploit the increased computational power to
speed up the~\ac{DDM} service~\cite{gda-simpat}. Therefore, an obvious
parallelization strategy for the intersection-finding problem is to
distribute the rectangles across the processor cores, so that each
core can work on a smaller problem. Interestingly, this approach fails
on all but the most trivial (and inefficient) algorithms.

In this paper we present a parallel implementation of~\ac{SBM} for
shared-memory processors. \ac{SBM} \cite{Raczy2005} is an efficient
solution to the $d$-dimensional rectangle intersection problem for the
special case $d=1$. Since any algorithm that can solve the
intersection problem in $d=1$ dimensions can be extended to $d > 1$
dimensions, \ac{SBM} is widely used to implement \ac{DDM}
services. Unfortunately, data dependencies in the \ac{SBM} algorithm
makes it difficult to exploit parallelism.\\

This paper is organized as follows. In Section~\ref{sec:related-work},
we review the state of the art concerning the~\ac{DDM} service. In
Section~\ref{sec:region-matching}, we describe some of the
existing~\ac{DDM} algorithms: brute force, grid-based, sequential
sort-based, and interval-tree matching. In
Section~\ref{sec:parallel-sort-matching}, we present the main
contribution of this work, i.e., a parallel version of the~\ac{SBM}
algorithm. In Section~\ref{sec:experimental-evaluation} we
experimentally evaluate the performance of parallel~\ac{SBM} on two
multicore processors. Finally, conclusions and future works will be
discussed in Section~\ref{sec:conclusions}.

\section{Related Work}\label{sec:related-work}

The matching part of~\ac{DDM} is a more specific instance of the
problem of identifying the intersecting pairs of (hyper)~rectangles
in a multidimensional metric space.

Data structures such as $k$-$d$ trees~\cite{Rosenberg1985} and
R-trees~\cite{Guttman1984} are able to efficiently store volumetric
objects and identify intersections. Such data structures are quite
complex to implement and, in many real-world situations, slower than
less efficient but simpler solutions~\cite{petty-1997}.  For example,
in~\cite{Devai2010} the authors introduced a rectangle-intersection
algorithm that is implemented using only simple data structures (i.e.,
arrays) and that can enumerate all $K$ intersections among $n$
rectangles with complexity $O(n \log n + K)$ time and $O(n)$ space.

Among the many matching algorithms that have been proposed for
enumerating all intersections among subscription and update extents,
the \ac{SBM}~\cite{Raczy2005} proved to be very efficient. \ac{SBM}
solves the region matching problem in one dimension; \ac{SBM} first
sorts the endpoints, and then scans the sorted set. In~\cite{Pan2011},
\ac{SBM} has been extended to deal with dynamic environments in which
extents are dynamic (both in terms of placement and size).  On the
other hand, \ac{SBM} has the drawback that it can not be trivially
parallelized due to the presence of a sequential scan phase that is
intrinsically serial. This is a serious limitation since the most of
modern processing architectures are multi or many-cores.

Only few parallel solutions for~\ac{DDM} and interest
matching~\cite{liu2013} have been proposed. Among them, the authors of
this paper have proposed the~\ac{ITM} algorithm for computing
intersections among $d$-rectangles~\cite{gda-dsrt-2013}.  \ac{ITM} is
based on an interval tree data structure, and after the tree is built,
exhibits an embarrassingly parallel structure.  The performance
evaluation reported in~\cite{gda-dsrt-2013} shows that the sequential
implementation of \ac{ITM} is competitive with the sequential
implementation of \ac{SBM}.

In~\cite{Yanbing2015}, a parallel ordered-relation-based matching algorithm
is proposed. The algorithm is composed of five phases: projection, sorting,
task decomposition, internal matching and external matching. In the 
experimental evaluation, a MATLAB implementation is compared with the
sequential \ac{SBM}. The results show that, with a high number of extents 
the proposed algorithm is faster than \ac{SBM}.

In~\cite{rao2013} the performance of parallel versions of \ac{BF} and 
grid-based matching (fixed, dynamic and hierarchical) are compared.
In this case, the preliminary results presented show that the parallel
\ac{BF} has a limited scalability and that, in this specific case,
the hierarchical grid-based matching has the best performance.

\section{The Region Matching Problem}\label{sec:region-matching}

In this section we define the~\ac{DDM} problem, and describe three
matching algorithms that have been thoroughly investigated in the
literature (brute-force, region-based and sort-based), in addition to
one that has been introduced recently (interval-tree matching).

Given two sets $\mathbf{S} = \{S_1, \ldots, S_n\}$ and
$\mathbf{U}=\{U_1, \ldots, U_m\}$ of $d$-dimensional rectangles with
sides parallel to the axes (called \emph{subscription extents} and
\emph{update extents}, respectively), the~\ac{DDM} problem consists of
identifying all intersections between a subscription extent and an
update extent. Formally, a \ac{DDM} algorithm must return the list of
all pairs $(S_i, U_j)$ such that $S_i \cap U_j \neq \emptyset$, $1
\leq i \leq n$, $1 \leq j \leq m$.

\begin{figure}[t]
\centering\includegraphics[scale=.7]{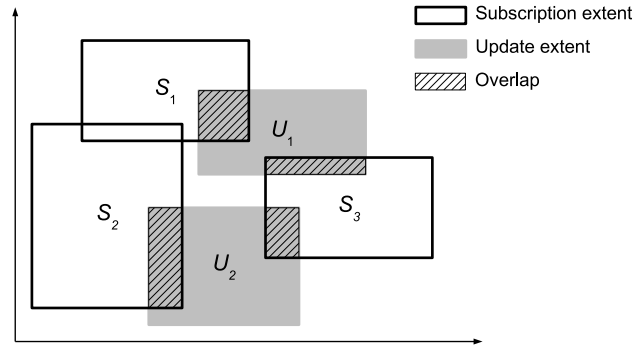}
\caption{An example of the Data Distribution Management problem in $d=2$ dimensions.}\label{fig:ddm_example}
\end{figure}

Figure~\ref{fig:ddm_example} shows an instance of the~\ac{DDM} problem
in $d=2$ dimensions with three subscription extents $\{S_1, S_2,
S_3\}$ and two update extents $\{U_1, U_2\}$. There are four overlaps
(intersections) between a subscription an update extent, namely $(S_1,
U_1)$, $(S_2, U_2)$, $(S_3, U_1)$, and~$(S_3, U_2)$. Note that $S_1$
and $S_2$ overlap, but this intersection is ignored since it involves
subscription extents only.

The time complexity of any~\ac{DDM} algorithm is output-sensitive,
since it depends on the size of the output in addition to the size of
the input. Therefore, every~\ac{DDM} algorithm that explicitly
enumerates all the~$K$ intersections requires time $\Omega(K)$. Since
there can be at most $n \times m$ intersections, the worst-case
complexity of the~\ac{DDM} problem is $O(n \times m)$.

\begin{algorithm}[t]
\caption{\textsc{Intersect-1D}$(x, y)$}\label{alg:intersect1d}
\begin{algorithmic}
\State \textbf{return} $x.\emph{low} \leq y.\emph{high} \wedge y.\emph{low} \leq x.\emph{high}$
\end{algorithmic}
\end{algorithm}

One of the key steps of any~\ac{DDM} algorithm is testing whether two
$d$-rectangles overlap. The special case $d=1$ is quite simple, as it
reduces to testing whether two closed intervals
$x=[x.\textit{low},x.\textit{high}]$, $y=[y.\textit{low},
  y.\textit{high}]$ intersect; this can be done in constant time: $x$
and $y$ overlap if and only if
\[
x.\emph{low} \leq y.\emph{high} \wedge y.\emph{low} \leq x.\emph{high}
\]
\noindent (see Algorithm~\ref{alg:intersect1d}).

The general case~$d>1$ can be reduced to the base case~$d=1$ by
observing that two $d$-rectangles overlap if and only if all their
projections along each dimension overlap. Therefore, we can invoke
Algorithm~\ref{alg:intersect1d} $d$ times, and compute the logical
``and'' of the results. Using this property, an algorithm that
enumerates all intersections among two sets of~$n$ and~$m$
one-dimensional segments in time~$O\left(f(n,m)\right)$ can be readily
extended to an~$O\left( d \times f(n,m) \right)$ algorithm for
reporting intersections among two sets of~$d$-rectangles. For this
reason, it is common practice in the~\ac{DDM} research community to
focus on the simpler one-dimensional case.

\subsection{Brute-Force Matching}

\begin{algorithm}[t]
\caption{\textsc{BruteForce-1D}$(\mathbf{S}, \mathbf{U})$}\label{alg:brute-force}
\begin{algorithmic}[1]
\State $n \gets |\mathbf{S}|$, $m \gets |\mathbf{U}|$, $L \gets \emptyset$
\For{$i \gets 1\ \textrm{to}\ n$}\label{alg:bf:for}
\For{$j \gets 1\ \textrm{to}\ m$}
\If {\Call{Intersect-1D}{$S_i, U_j$}}
\State $L \gets L \cup (S_i, U_j)$
\EndIf
\EndFor
\EndFor\label{alg:bf:endfor}
\State \textbf{return} $L$
\end{algorithmic}
\end{algorithm}

The simplest solution to the $1$-dimensional segment intersection
problem is the~\ac{BF} approach, also called Region-Based matching
(Algorithm~\ref{alg:brute-force}). The~\ac{BF} algorithm, as the name
suggests, checks all $n \times m$ subscription-update pairs and
inserts every intersection into a list~$L$.

Despite its simplicity, the~\ac{BF} algorithm is extremely inefficient
since it requires time~$O(nm)$. However, it exhibits an
embarrassingly parallel structure since the loop iterations
(lines~\ref{alg:bf:for}--\ref{alg:bf:endfor}) are independent. This
makes parallelization of the the~\ac{BF} algorithm trivial; when~$P$
processors are available, the amount of work performed by each
processor is~$O\left(nm / P\right)$.

\subsection{Grid-Based Matching}\label{sec:grid-based-matching}

\begin{figure}[t]
\centering\includegraphics[scale=.7]{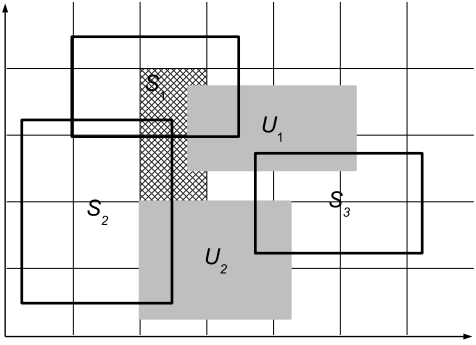}
\caption{Grid-based matching in $d=2$ dimensions.}\label{fig:ddm_example_grid}
\end{figure}

The~\acf{GB} matching algorithm proposed by Boukerche and
Dzermajko~\cite{Boukerche2001} improves over~\ac{BF} matching.
\ac{GB} works by partitioning the routing space into a regular mesh of
$d$-dimensional cells. Each subscription or update extent is mapped to
the grid cells it overlaps with. Events generated by an update
extent~$U_j$ are sent to all subscription extents that share at least
one cell with~$U_j$. A filtering mechanism must then be applied to
avoid delivering of spurious events. For example, in
Figure~\ref{fig:ddm_example_grid} we see that~$S_2$ shares the hatched
grid cells with~$U_1$, but does not overlap with~$U_1$. Hence,
the~\ac{GB} matching algorithm would send notifications from $U_1$ to
$S_2$ that will need to be filtered out.

A simple filtering mechanism consists on the application of
the~\ac{BF} algorithm to each grid cell. If the routing space is
partitioned into~$G$ cells and all extents are evenly distributed,
each cell will overlap with $n/G$ subscription and $m/G$ update
extents on average. Therefore, the brute force approach applied to
each cell will require~$O(nm / G^2)$ operations; since there are~$G$
cells, the overall worst-case complexity of~\ac{GB} matching is~$O(nm
/ G)$. Therefore, in the ideal case~\ac{GB} can decrease the matching
complexity by a factor~$G$ with respect to~\ac{BF}. Unfortunately,
when cells are small (and therefore~$G$ is large) each extent is
mapped to a larger number of cells, which increases the computation
time.

\subsection{Interval-Tree Matching}\label{sec:interval-tree}

The \acf{ITM} algorithm~\cite{gda-dsrt-2013} is based on the
\emph{interval tree} data structure that solves the matching problem
in one dimension. An interval tree is a balanced search tree that
stores a dynamic set of intervals, supporting insertions, deletions,
and queries to get the list of segments intersecting a given
interval~$q$.  Different implementations of interval trees are
possible, depending on the structure of the underlying search tree;
the implementation described in~\cite{gda-dsrt-2013} is based on~AVL
trees~\cite{avl}.

\begin{figure}[t]
\centering\includegraphics[width=\columnwidth]{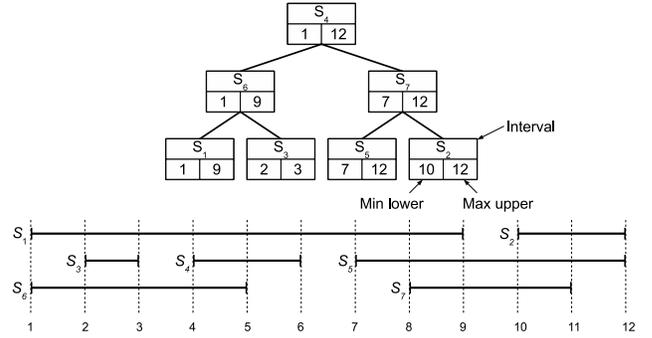}
\caption{A set of intervals and the corresponding interval tree.}\label{fig:interval_tree}
\end{figure}

Each node~$x$ of the~AVL tree holds three fields: (\emph{i})~an
interval $x.\textit{in}$, represented by its lower and upper bounds;
(\emph{ii})~the minimum lower bound~$x.\textit{minlower}$ among all
intervals stored at the subtree rooted at~$x$; (\emph{iii})~the
maximum upper bound~$x.\textit{maxupper}$ among all intervals stored
at the subtree rooted at~$x$. Nodes are kept sorted according to the
interval lower bounds. Figure~\ref{fig:interval_tree} shows a set of
intervals and the corresponding interval tree representation.

Insertions and deletions are handled according to the normal rules
for~AVL trees, with the additional requirement that any update of the
values of~$\textit{maxupper}$ and~$\textit{minlower}$ must be
propagated up to the tree root. Since the height of an~AVL tree
is~$O(\log n)$, insertions and deletions in the augmented data
structure require~$O(\log n)$ time in the worst case. The storage
requirement is~$O(n)$.

\begin{algorithm}[t]
\caption{\textsc{Interval-Tree-Matching-1D}$(\mathbf{S}, \mathbf{U})$}\label{alg:int-tree-matching}
\begin{algorithmic}[1]
\Function{Interval-Query}{$x, q$}

\State $L \gets \emptyset$
\If{$x = \textit{null}$ \textbf{or} $x.\textit{maxupper} < q.\textit{lower}$ \textbf{or} $x.\textit{minlower} > q.\textit{upper}$}
\State \textbf{return} $L$
\EndIf

\State $L \gets $ \Call{Interval-Query}{$x.\textit{left}, q$}

\If{\Call{Intersect-1D}{$x.\textit{in}, q$}}
\State $L \gets L \cup \{(x.\textit{in}, q)\}$
\EndIf

\If{$q.\textit{upper} \geq x.\textit{in}.\textit{lower}$}
\State $L \gets L \cup \mathrm{}$ \Call{Interval-Query}{$x.\textit{right}, q$}
\EndIf

\State \textbf{return} $L$
\EndFunction

\Statex

\State $n \gets |\mathbf{S}|$, $m \gets |\mathbf{U}|$, $L \gets \emptyset$
\State $T \gets \textrm{create interval tree for $\mathbf{S}$}$\label{alg:int-tree-create}
\For{$j \gets 1~\mathrm{to}~m$}\label{alg:int-tree-loop-begin}
\State $L \leftarrow L \cup \mathrm{}$ \Call{Interval-Query}{$T.\textit{root}, U_j$}
\EndFor\label{alg:int-tree-loop-end}
\State \textbf{return} $L$
\end{algorithmic}
\end{algorithm}

Function~\textsc{IntTree-Matching-1D}
(Algorithm~\ref{alg:int-tree-matching}) returns the list of
intersections among the set~$\mathbf{S}$ of subscription intervals and
the set~$\mathbf{U}$ of update intervals. This is done by first
building an interval tree~$T$ containing all elements in~$\mathbf{S}$
(line~\ref{alg:int-tree-create}); then, for each update interval $U_j
\in \mathbf{U}$, the algorithm calls function
$\textsc{Interval-Query}(x, q)$ to identify all subscriptions that
intersect~$U_j$
(lines~\ref{alg:int-tree-loop-begin}--\ref{alg:int-tree-loop-end}).
The function returns the list of intersections of the update
interval~$q$ with the segments stored in the subtree rooted at~$x$
($T.\textit{root}$ is the root
of~$T$). Function~\textsc{Interval-Query} performs a visit of the
interval tree data structure, using the values of
attributes~$x.\textit{minlower}$ and~$x.\textit{maxupper}$ of each
node~$x$ to steer the visit out of the subtrees that would yield no
matches.

An interval tree can be created in time~$O(n \log n)$; the total query
time is $O\left( \min\{mn, (K+1) \log n\}\right)$, $K \leq nm$ being
the number of intersections involving all subscription and all update
intervals~\cite{gda-dsrt-2013}.  When executed on a shared-memory
multiprocessor with~$P$ cores, the iterations of the~\textbf{for} loop
in Algorithm~\ref{alg:int-tree-matching},
lines~\ref{alg:int-tree-loop-begin}--\ref{alg:int-tree-loop-end} can
be split across the cores, with the provision that updates to the
result list $L$ are serialized. The only remaining serial part is the
construction of the interval tree; while concurrent balanced search
trees have been proposed in the literature~\cite{Medidi98, Park01} it
is unclear whether they can be used as drop-in replacements.

\begin{algorithm}[t]
\caption{\textsc{Sort-Based-Matching-1D}$(\mathbf{S}, \mathbf{U})$}\label{alg:sbm}
\begin{algorithmic}[1]
\State $L \gets \emptyset$, $T \gets \emptyset$
\ForAll{extents $x \in \mathbf{S} \cup \mathbf{U}$}
\State Insert $x.\textit{lower}$ and $x.\textit{upper}$ in $T$
\EndFor
\State Sort $T$ in non-decreasing order
\State $\texttt{SubSet} \gets \emptyset$, $\texttt{UpdSet} \gets \emptyset$
\ForAll{points $t \in T$ in non-decreasing order}\label{alg:sbm-loop-start}
\If{$t$ belongs to subscription extent $S_i$}
\If{$t$ is the lower bound of $S_i$}
\State $\texttt{SubSet} \gets \texttt{SubSet} \cup \{S_i\}$
\Else
\State $\texttt{SubSet} \gets \texttt{SubSet} \setminus \{S_i\}$
\ForAll{$x \in \texttt{UpdSet}$}
\State $L \gets L \cup \{(S_i, x)\}$
\EndFor
\EndIf
\Else\Comment{$t$ belongs to update extent $U_j$}
\If{$t$ is the lower bound of $U_j$}
\State $\texttt{UpdSet} \gets \texttt{UpdSet} \cup \{U_j\}$
\Else
\State $\texttt{UpdSet} \gets \texttt{UpdSet} \setminus \{U_j\}$
\ForAll{$x \in \texttt{SubSet}$}
\State $L \gets L \cup \{(x, U_j)\}$
\EndFor
\EndIf
\EndIf
\EndFor\label{alg:sbm-loop-end}
\State \textbf{return} $L$
\end{algorithmic}
\end{algorithm}

\begin{figure}[t]
\centerline{\includegraphics[width=1.1\columnwidth]{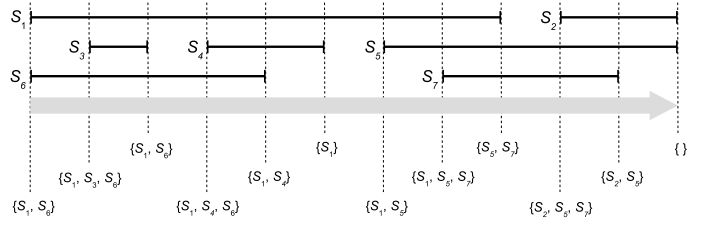}}
\caption{Value assigned by the~\ac{SBM} algorithm to the
  \texttt{SubSet} variable as the endpoints are swept from left to
  right.}\label{fig:sbm-example}
\end{figure}

\subsection{Sort-Based Matching}\label{sec:sort-based}

The~\acl{SBM} algorithm~\cite{Jun2002,Raczy2005} is an efficient
solution to the~\ac{DDM} problem. Algorithm~\ref{alg:sbm} illustrates
\ac{SBM} in its basic form: given a set $\mathbf{S}$ of $n$
subscription intervals, and a set $\mathbf{U}$ of $m$ update
intervals, the algorithm considers each of the $2 \times (n+m)$
endpoints in non-decreasing order; two sets \texttt{SubSet} and
\texttt{UpdSet} are used to keep track of the active subscription and
update intervals at every point~$t$; we say that an interval is active
at $t$ if its lower endpoint has time $\leq t$, and its upper endpoint
has time $> t$. For example, Figure~\ref{fig:sbm-example} shows the
values of \texttt{SubSet} while the~\ac{SBM} sweeps through a set of
subscription intervals (update intervals are handled in exactly the
same way). When the upper bound of an interval is encountered, the
list of intersections~$L$ is updated accordingly.

Let $N=n+m$ be the total number of endpoints; then, the~\ac{SBM}
algorithm uses simple data structures and requires $O\left( N \log N
\right)$ time to sort the vector of endpoints, plus~$O(N)$ time to
scan the sorted vector. During the scan phase, $O(K)$ time is spent in
total to transfer the information from the sets \texttt{SubSet} and
\texttt{UpdSet} to the intersection list $L$. The overall
computational cost of~\ac{SBM} is $O\left( N \log N + K \right)$ ($K$
is the number of intersections).

\section{Parallel Sort-based Matching}\label{sec:parallel-sort-matching}

In this section we describe a parallel version of the~\ac{SBM}
algorithm, using Algorithm~\ref{alg:sbm} as the starting point.

We have seen that \ac{SBM} operates in two phases: first, the list $T$
of endpoints is sorted; then, the sorted list is traversed to compute
the values of the~\texttt{SubSet} and~\texttt{UpdSet} variables, from
which the list of overlaps is derived. On a shared-memory architecture
with~$P$ processors, the sorting phase can be realized using a
parallel sorting algorithm~\cite{Wheat92,Cole88}. The traversal of the
sorted list of endpoints (Algorithm~\ref{alg:sbm}
lines~\ref{alg:sbm-loop-start}--\ref{alg:sbm-loop-end}) is, however,
more challenging to execute in parallel. Ideally, we would like to
split the list $T$ into~$P$ segments of equal size $T_0, \ldots,
T_{P-1}$, and assign each segment to a processor. Unfortunately, this
is made difficult by the loop-carried dependencies caused by the
variables \texttt{SubSet} and \texttt{UpdSet}, whose values are
modified at each iteration.

\begin{algorithm}[t]
\caption{\textsc{Parallel-SBM-1D}$(\mathbf{S}, \mathbf{U})$}\label{alg:par-sbm}
\begin{algorithmic}[1]
\State $L \gets \emptyset$, $T \gets \emptyset$
\ForAll{extents $x \in \mathbf{S} \cup \mathbf{U}$ \textbf{in parallel}}
\State Insert $x.\textit{lower}$ and $x.\textit{upper}$ in $T$
\EndFor
\State Sort $T$ \textbf{in parallel}, in non-decreasing order
\State Split $T$ into $P$ segments $T_0, \ldots, T_{P-1}$
\State $\langle$Initialize $\texttt{SubSet}[0..P-1]$ and $\texttt{UpdSet}[0..P-1] \rangle$\label{alg:par-sbm-init}
\For{$p \gets 0~\textrm{to}~P-1$ \textbf{in parallel}}
\ForAll{endpoints $t \in T_p$ in non-decreasing order}
\If{$t$ belongs to subscription extent $S_i$}
\If{$t$ is the lower bound of $S_i$}
\State $\texttt{SubSet}[p] \gets \texttt{SubSet}[p] \cup \{S_i\}$
\Else
\State $\texttt{SubSet}[p] \gets \texttt{SubSet}[p] \setminus \{S_i\}$
\ForAll{$x \in \texttt{UpdSet}[p]$}
\State \textbf{Atomic} $L \gets L \cup \{(S_i, x)\}$
\EndFor
\EndIf
\Else\Comment{$t$ belongs to update extent $U_j$}
\If{$t$ is the lower bound of $U_j$}
\State $\texttt{UpdSet}[p] \gets \texttt{UpdSet}[p] \cup \{U_j\}$
\Else
\State $\texttt{UpdSet}[p] \gets \texttt{UpdSet}[p] \setminus \{U_j\}$
\ForAll{$x \in \texttt{SubSet}[p]$}
\State \textbf{Atomic} $L \gets L \cup \{(x, U_j)\}$
\EndFor
\EndIf
\EndIf
\EndFor
\EndFor
\State \textbf{return} $L$
\end{algorithmic}
\end{algorithm}

Let us pretend that the scan phase can be parallelized somehow. Then,
a parallel version of~\ac{SBM} would look like
Algorithm~\ref{alg:par-sbm} (line~\ref{alg:par-sbm-init} will be
explained shortly). The major difference between
Algorithm~\ref{alg:par-sbm} and its sequential counterpart is that the
former uses two arrays $\texttt{SubSet}[p]$ and $\texttt{UpdSet}[p]$
instead of the scalar variables \texttt{SubSet} and
\texttt{UpdSet}. This allows each core to operate on its private copy
of the subscription and update sets, achieving the maximum level of
parallelism.

It is not difficult to see that Algorithm~\ref{alg:par-sbm} is
equivalent to the sequential~\ac{SBM} (i.e., they return the same
result) if and only if $\texttt{SubSet}[0..P-1]$ and
$\texttt{UpdSet}[0..P-1]$ are properly initialized. Specifically,
$\texttt{SubSet}[p]$ and $\texttt{UpdSet}[p]$ must be initialized with
the values that the sequential~\ac{SBM} algorithm assigns to
\texttt{SubSet} and \texttt{UpdSet} right after the last endpoint of
$T_{p-1}$ is processed, $p=1, \ldots, P-1$; $\texttt{SubSet}[0]$ and
$\texttt{UpdSet}[0]$ must be initialized to the empty set.

It turns out that the content of the arrays $\texttt{SubSet}[0..P-1]$
and $\texttt{UpdSet}[0..P-1]$ can be computed efficiently using a
\emph{prefix computation} (also called \emph{scan} or
\emph{prefix-sum}). To make this paper self-contained, we provide
details on prefix computations before illustrating the missing part of
the parallel~\ac{SBM} algorithm.

\paragraph*{Prefix computations}
A prefix computation consists of a sequence of $N > 0$ data items
$x_0, \ldots, x_{N-1}$ and an associative operator~$\oplus$. There are
two types of prefix computations: the \emph{inclusive scan} operation
produces a new sequence of $N$ data items $y_0, \ldots, y_{N-1}$ such
that:

\begin{alignat*}{2}
y_0 &= x_0 \\
y_1 &= y_0 \oplus x_1 &&= x_0 \oplus x_1 \\
y_2 &= y_1 \oplus x_2 &&= x_0 \oplus x_1 \oplus x_2 \\
&\vdots \\
y_{N-1} &= y_{N-2} \oplus x_{N-1} &&= x_0 \oplus x_1 \oplus \ldots \oplus x_{N-1}
\end{alignat*}

\noindent while the \emph{exclusive scan} operation produces the
sequence $z_0, z_1, \ldots z_{N-1}$ such that:

\begin{alignat*}{2}
z_0 &= 0 \\
z_1 &= z_0 \oplus x_0 &&= x_0 \\
z_2 &= z_1 \oplus x_1 &&= x_0 \oplus x_1 \\
&\vdots \\
z_{N-1} &=  z_{N-2} \oplus x_{N-2} &&= x_0 \oplus x_1 \oplus \ldots \oplus x_{N-2}
\end{alignat*}

\noindent where $0$ is the neutral element of operator $\oplus$, i.e.,
$0 \oplus x = x$.

Blelloch~\cite{Blelloch89} showed that the prefix sums of~$N$ items
can be computed in time $O(N/P + \log P)$ using $P < N$ processors on
a shared-memory multiprocessor by organizing the computation as a
tree. The $O(N/P + \log P)$ time is optimal when $N/P > \log P$.  In
our algorithm we use a simpler two-level mechanism that achieves
running time $O(N/P + P)$, which is still optimal when $N/P > P$.
This is usually the case, since the current generation of CPUs have a
small number of cores (e.g., $P \leq 72$ for the Intel Xeon Phi) and
the number of extents $N$ is usually very large.  We remark that the
parallel~\ac{SBM} algorithm can be readily implemented with the
tree-structured reduction operation, and therefore will still be
competitive on future generations of processors with a higher number
of cores.

\begin{figure}[t]
\centering\includegraphics[scale=.7]{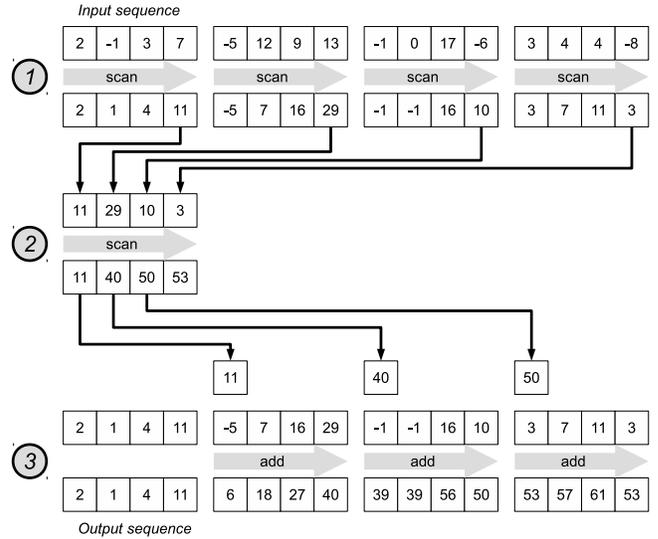}
\caption{Parallel prefix sum computation.}\label{fig:scan}
\end{figure}

Figure~\ref{fig:scan} illustrates an example of parallel (inclusive)
scan with $P=4$ processors, assuming that the $\oplus$ operator is the
numeric addition. The computation involves two parallel steps, and one
serial step which is executed by a single processor that we call the
master. \circlenum{1}~The input sequence is splitted across the
processors, and each processor computes the prefix sum of the elements
in its portion. \circlenum{2}~The master computes the prefix sum of
the $P$ last local sums. \circlenum{3}~The master scatters the
\emph{first} $(P-1)$ computed values (prefix sums of the last local
sums) to the \emph{last} $(P-1)$ processors. Each processor, except
the first one, adds (more precisely, applies the $\oplus$ operator)
the received value to the prefix sums from step~\circlenum{1},
producing a portion of the output sequence. Steps~\circlenum{1}
and~\circlenum{3} require time $O(N/P)$, while step~\circlenum{2} is
executed by the master only in time $O(P)$, yielding a total cost of
$O(N/P + P)$.

\begin{algorithm}[h]
\caption{$\langle$Initialize $\texttt{SubSet}[0..P\!-\!1]$ and $\texttt{UpdSet}[0..P\!-\!1] \rangle$}\label{alg:par-sbm-cont}
\begin{algorithmic}[1]
\Statex\Comment{Executed by all cores in parallel}
\For{$p \gets 0~\textrm{to}~P-1$ \textbf{in parallel}}\label{alg:par-sbm-loop-begin}
\State $\texttt{Sadd}[p] \gets \emptyset$, $\texttt{Sdel}[p] \gets \emptyset$, $\texttt{Uadd}[p] \gets \emptyset$, $\texttt{Udel}[p] \gets \emptyset$
\ForAll{points $t \in T_p$ in non-decreasing order}
\If{$t$ belongs to subscription extent $S_i$}
\If{$t$ is the lower bound of $S_i$}
\State $\texttt{Sadd}[p] \gets \texttt{Sadd}[p] \cup \{S_i\}$
\ElsIf{$S_i \in \texttt{Sadd}[p]$}
\State $\texttt{Sadd}[p] \gets \texttt{Sadd}[p] \setminus \{S_i\}$
\Else
\State $\texttt{Sdel}[p] \gets \texttt{Sdel}[p] \cup \{S_i\}$
\EndIf
\Else\Comment{$t$ belongs to update extent $U_j$}
\If{$t$ is the lower bound of $U_j$}
\State $\texttt{Uadd}[p] \gets \texttt{Uadd}[p] \cup \{U_j\}$
\ElsIf{$U_j \in \texttt{Uadd}[p]$}
\State $\texttt{Uadd}[p] \gets \texttt{Uadd}[p] \setminus \{U_j\}$
\Else
\State $\texttt{Udel}[p] \gets \texttt{Udel}[p] \cup \{U_j\}$
\EndIf
\EndIf
\EndFor
\EndFor\label{alg:par-sbm-loop-end}
\Statex\Comment{Executed by the master only}
\State $\texttt{SubSet}[0] \gets \emptyset$, $\texttt{UpdSet}[0] \gets \emptyset$\label{alg:par-sbm-step-2-begin}
\For{$p \gets 1~\textrm{to}~P-1$}
\State $\texttt{SubSet}[p] \gets \texttt{SubSet}[p-1] \cup \texttt{Sadd}[p-1] \setminus \texttt{Sdel}[p-1]$
\State $\texttt{UpdSet}[p] \gets \texttt{UpdSet}[p-1] \cup \texttt{Uadd}[p-1] \setminus \texttt{Udel}[p-1]$
\EndFor\label{alg:par-sbm-step-2-end}
\end{algorithmic}
\end{algorithm}

\begin{figure*}[t]
\centering\includegraphics[scale=.7]{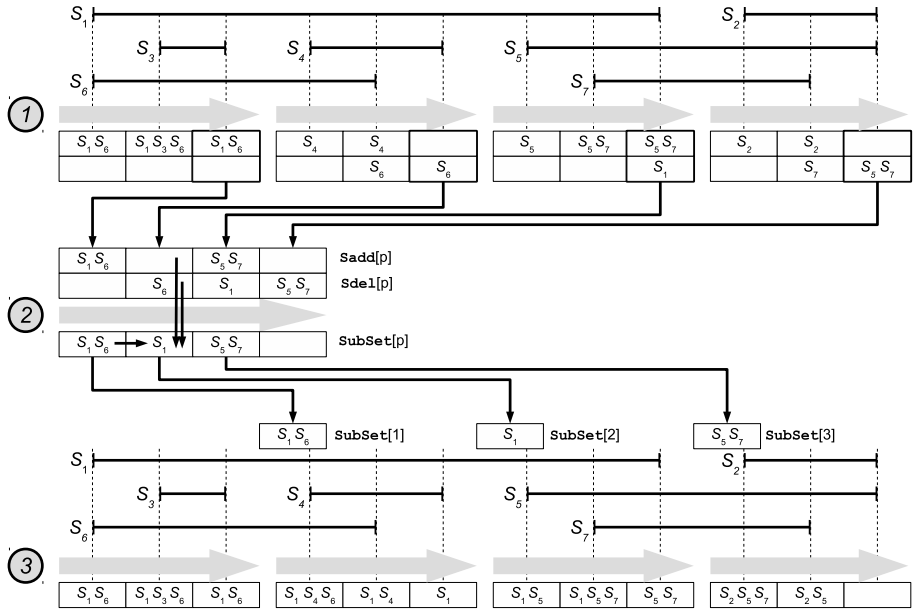}
\caption{Parallel prefix computation for the~\ac{SBM} algorithm.}\label{fig:par-sbm}
\end{figure*}

\paragraph*{Parallel Sort Matching}
We are now ready to complete the description of the parallel~\ac{SBM}
algorithm by showing how to fill the arrays $\texttt{SubSet}[p]$ and
$\texttt{UpdSet}[p]$ in parallel. To better illustrate the steps
involved, we refer to the example in Figure~\ref{fig:par-sbm}. In the
figure, we consider subscription extents only, since the procedure for
update extents is the same.

The sorted list of endpoints $T$ is evenly split into $P$ segments
$T_0, \ldots, T_{P-1}$. Processor $p$ scans the endpoints $t \in T_p$
in non-decreasing order, updating four auxiliary variables
$\texttt{Sadd}[p]$, $\texttt{Sdel}[p]$, $\texttt{Uadd}[p]$, and
$\texttt{Udel}[p]$. Informally, $\texttt{Sadd}[p]$ and
$\texttt{Sdel}[p]$ (resp. $\texttt{Uadd}[p]$ and $\texttt{Udel}[p]$)
contain the endpoints that the sequential~\ac{SBM} algorithm would
add/remove from \texttt{SubSet} (resp. \texttt{UpdSet}) while scanning
the endpoints belonging to segment $T_p$. More formally, at the
end of each local scan the following invariants hold:
\begin{enumerate}
\item $\texttt{Sadd}[p]$ (resp. $\texttt{Uadd}[p]$) contains the
  subscription (resp. update) intervals whose lower endpoint belongs
  to $T_p$, and whose upper endpoint does not belong to $T_p$;
\item $\texttt{Sdel}[p]$ (resp. $\texttt{Udel}[p]$) contains the
  subscription (resp. update) intervals whose upper endpoint belongs
  to $T_p$, and whose lower endpoint does not belong to $T_p$.
\end{enumerate}
This step is realized by
lines~\ref{alg:par-sbm-loop-begin}--\ref{alg:par-sbm-loop-end} of
Algorithm~\ref{alg:par-sbm-cont}, and its effects are shown in
Figure~\ref{fig:par-sbm}~\circlenum{1}. The figure reports the values
of $\texttt{Sadd}[p]$ and $\texttt{Sdel}[p]$ after each endpoint has
been processed; the algorithm does not store every intermediate value,
since only the last ones (within thick boxes) will be needed by the
next step.

Once all $\texttt{Sadd}[p]$ and $\texttt{Sdel}[p]$ are available, the
next step is executed by the master and consists of computing the
values of $\texttt{SubSet}[p]$ and $\texttt{UpdSet}[p]$, $p=0, \ldots,
P-1$. Recall from the discussion above that $\texttt{SubSet}[p]$
(resp. $\texttt{UpdSet}[p]$) is the set of active subscription
(resp. update) intervals that would be identified by the
sequential~\ac{SBM} algorithm right after the end of segment $T_0 \cup
\ldots \cup T_{p-1}$. The values of $\texttt{SubSet}[p]$ and
$\texttt{SubSet}[p]$ are related to $\texttt{Sadd}[p]$,
$\texttt{Sdel}[p]$, $\texttt{Uadd}[p]$ and $\texttt{Udel}[p]$ as
follows:
\begin{align*}
\texttt{SubSet}[p] &= \begin{cases}
\emptyset & \mbox{if $p=0$} \\
\texttt{SubSet}[p-1] \cup \texttt{Sadd}[p] \setminus \texttt{Sdel}[p] & \mbox{if $p>0$}
\end{cases} \\
\texttt{UpdSet}[p] &= \begin{cases}
\emptyset & \mbox{if $p=0$} \\
\texttt{UpdSet}[p-1] \cup \texttt{Uadd}[p] \setminus \texttt{Udel}[p] & \mbox{if $p>0$}
\end{cases}
\end{align*}
Intuitively, the set of active intervals at the end of $T_p$ can be
computed from those active at the end of $T_{p-1}$, plus the intervals
that became active in $T_p$, minus those that ceased to be active in
$T_p$.

Lines~\ref{alg:par-sbm-step-2-begin}--\ref{alg:par-sbm-step-2-end} of
Algorithm~\ref{alg:par-sbm-cont} take care of this computation; see
also Figure~\ref{fig:par-sbm}~\circlenum{2} for an example. Once the
initial values of $\texttt{SubSet}[p]$ and $\texttt{UpdSet}[p]$ have
been computed, Algorithm~\ref{alg:par-sbm} can be resumed to identify
the list of overlaps.

\paragraph*{Asymptotic Analysis} 
We now analyze the asymptotic cost of parallel \ac{SBM}.
Algorithm~\ref{alg:par-sbm} consists of three phases:
\begin{enumerate}
\item Fill the array of endpoints $T$, and sort $T$ in non-decreasing
  order; if $P$ processors are available, this step requires total
  time $O\left(N \log N / P\right)$, where $N$ is the total number of
  subscription and update extents, using a suitable sorting algorithm
  such as parallel merge sort~\cite{Cole88}.
\item Compute the initial values of $\texttt{SubSet}[p]$ and
  $\texttt{UpdSet}[p]$, for each $p=0, \ldots, P-1$; this phase
  requires $O\left(N/P + P\right)$ steps using the two-level scan
  shown on Algorithm~\ref{alg:par-sbm-cont}; the time can be further
  reduced to $O\left(N/P + \log P\right)$ steps using a
  tree-structured reduction~\cite{Blelloch89}.
\item Perform the final local scans. Each scan can be completed in
  $O(N/P)$ steps.
\end{enumerate}

Note, however, that phases~2 and~3 require the manipulation of data
structures to hold sets of endpoints, supporting insertions and
removals of single elements and whole sets. Therefore, a single step
of the algorithm has a non-constant time complexity that depends on
the actual implementation of sets and the number of elements they
contain. Furthermore, during phase~3 total time $O(K)$ is spend
cumulatively by all processors to push all $K$ intersections into the
result list~$L$.

\section{Experimental Evaluation}\label{sec:experimental-evaluation}

In this section we evaluate the performance and scalability of
parallel~\ac{SBM} with respect to parallel versions of the~\ac{BF}
and~\ac{ITM} algorithms. \ac{BF} and \ac{ITM} are considered because
both exhibit an embarrassingly parallel structure, and \ac{ITM} has
already been shown to be more computationally efficient than
\ac{BF}~\cite{gda-dsrt-2013}. In the present study we do not consider
the~\ac{GB} algorithm: while it can be very fast and contains easily
exploitable parallelism, its efficiency depends on the grid size~$G$
that should either be judiciously selected, or adaptively defined by
means of non-trivial heuristics~\cite{Boukerche2002}. Therefore, to
reduce the number of degrees of freedom we restrict our study to
algorithms that have no tunable parameters, postponing a more complete
study to a forthcoming paper.
To foster the reproducibility of our experiments, all the source code
used in this performance evaluation, and the raw data obtained in the
experiments execution, are freely available on the
Web\footnote{\url{http://pads.cs.unibo.it}}.

The~\ac{BF} and~\ac{ITM} algorithms have been implemented in C, and
the parallel~\ac{SBM} algorithm has been implemented in C++. We used
the GNU C Compiler (GCC) version 4.8.4 with the
\verb+-O3 -fopenmp -D_GLIBCXX_PARALLEL+ flags to turn on optimization
and to enable parallel constructs at the compiler and library
levels. Specifically, the \verb+-fopenmp+ flag allows the compiler to
process OpenMP directives in the source code~\cite{OpenMP}. OpenMP is
an open interface supporting shared memory parallelism in the C, C++
and FORTRAN programming languages. OpenMP allows the programmer to
label specific sections of the source code as parallel regions; the
compiler takes care of dispatching portions of these regions to
separate threads, that the~\ac{OS} can schedule on separate processors
or cores. In the C/C++ languages, OpenMP directives are specified
using \verb+#pragma+ compiler hints. The OpenMP standard also defines
a set of library functions that can be called by the programmer to
query and control the execution environment programmatically.

Both the~\ac{BF} and~\ac{ITM} algorithms required a single
\verb+omp parallel for+ directive to parallelize their inner loop. The
parallel~\ac{SBM} algorithm was more complex, and its implementation
benefited from the use of some of the data structures and algorithms
provided by the C++~\ac{STL}~\cite{Stroustrup13}. Specifically, to
sort the endpoints we used the parallel \verb+std::sort+ function
provided by the~\ac{STL} extensions for
parallelism~\cite{parallel-stl}. Indeed, the GNU~\ac{STL} provides
several parallel sort algorithms (multiway mergesort and quicksort
with various splitting heuristics) that are automatically selected at
compile time when the \verb+-D_GLIBCXX_PARALLEL+ compiler flag is
given. The remaining part of the \ac{SBM} algorithm has been
parallelized using explicit OpenMP directives.

The~\acf{SBM} algorithm requires a suitable data structure to store
the sets of endpoints \texttt{SubSet} and \texttt{UpdSet} (see
Algorithms~\ref{alg:par-sbm} and~\ref{alg:par-sbm-cont}).
Parallel~\ac{SBM} puts a higher strain on this data structure with
respect to its sequential counterpart, since it requires efficient
support for unions and differences between sets, in addition to
insertions and deletions of single elements. We have experimented with
three implementations for sets: (\emph{i})~bit vectors based on the
\verb+std::vector<bool>+ \ac{STL} container (note that
\verb+std::bitset+ can not be used, since it requires the set size to
be known at compile time); (\emph{ii})~an ad-hoc implementation of bit
vectors based on raw memory manipulation; (\emph{iii})~the
\verb+std::set+ container, which in the case of the GNU \ac{STL} is
based on Red-Black trees~\cite{Bayer1972}.  The latter turned out to
be the most efficient, so the performance results reported in this
section refer to the \verb+std::set+ container.

\begin{table}[ht]
\centering\begin{tabular}{lll}
\toprule
 & \texttt{solaris} & \texttt{titan} \\
\midrule
CPU             & Intel Xeon    & Intel Core    \\
                & E5-2640       & i7-5820K      \\
Clock frequency & 2.00 GHz      & 3.30 GHz      \\
Processors      & 2             & 1             \\
Total cores     & 16            & 6             \\ 
HyperThreading  & Yes           & Yes           \\
RAM             & 128 GB        & 64 GB         \\
L3 cache size   & 20480 KB      & 15360 KB      \\
\bottomrule
\end{tabular}
\caption{Hardware specifications of the machines used for the
  experimental evaluation.}\label{tab:spec}
\end{table}

\begin{figure*}[ht]
  \centering%
  \subfigure[Wall-clock time\label{fig:wct-1M}]{\includegraphics[width=.95\columnwidth]{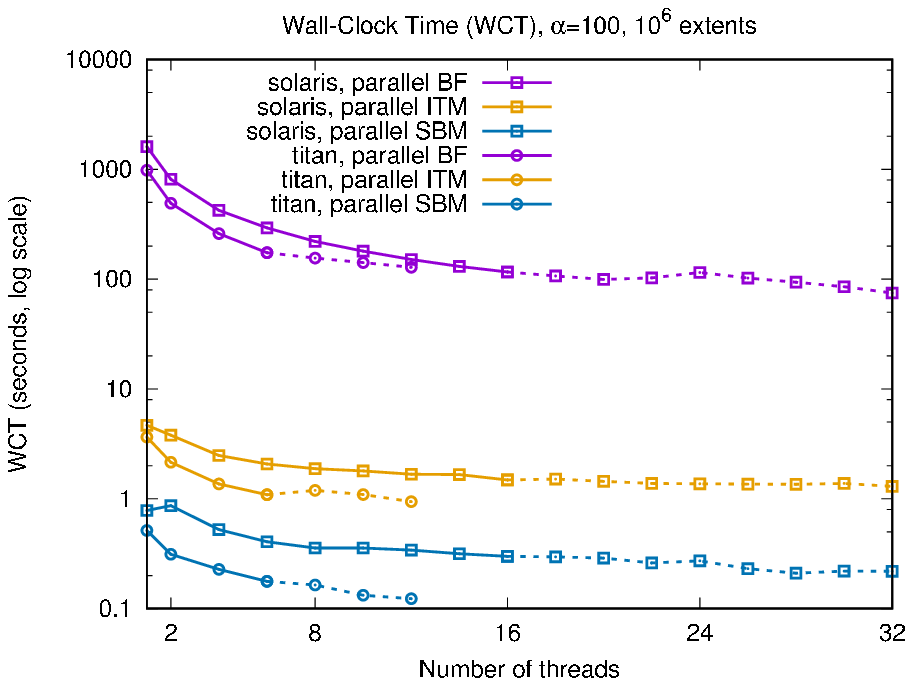}}\quad
  \subfigure[Speedup\label{fig:speedup-1M}]{\includegraphics[width=.95\columnwidth]{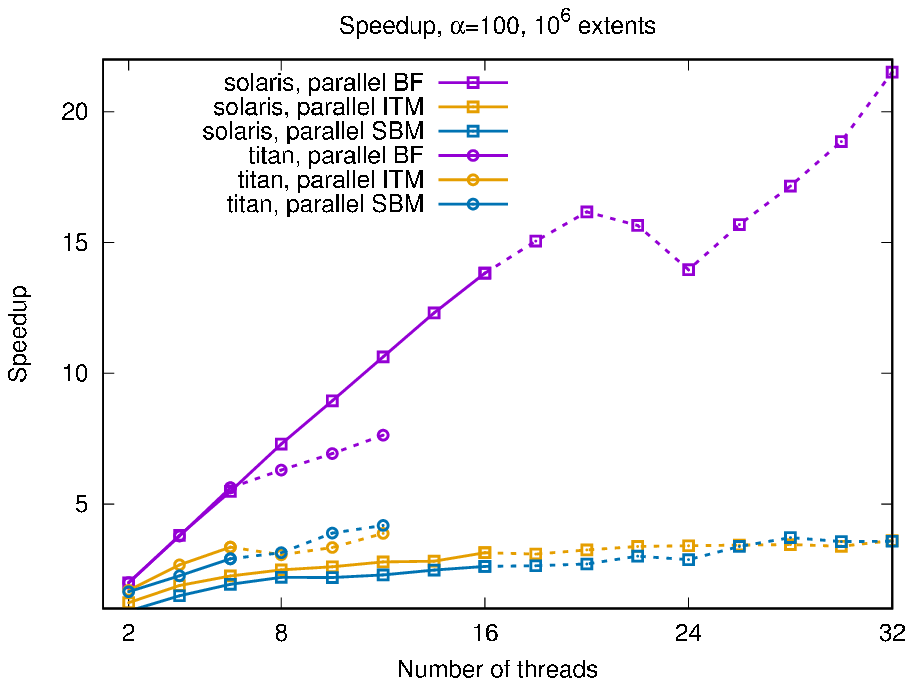}}
  \caption{Wall clock time and speedup of
    parallel~\{\ac{BF},~\ac{ITM},~\ac{SBM}\} with $N=10^6$ extents and
    overlapping degree $\alpha=100$. Dashed lines indicate the region
    where the number of OpenMP threads exceeds the number of CPU
    cores, and therefore \ac{HT} comes into
    play.}\label{fig:wct-speedup-1M}
\end{figure*}

\paragraph*{Experimental setup}
The experiments have been carried out on two different machines,
called \texttt{solaris} and \texttt{titan}, both running the 64 bit
version of the Ubuntu 14.04.05 LTS \ac{OS}. The hardware
specifications are reported in Table~\ref{tab:spec}: \texttt{solaris}
has two Intel Xeon processors with 8 cores each (16 cores total);
\texttt{titan} has a single Intel Core i7 processor with 6 cores. Both
types of processors employ the~\ac{HT} technology~\cite{HT}. In
\ac{HT}-enabled CPUs some functional components are duplicated, but
there is a single main execution unit for physical core. From the
point of view of the~\ac{OS}, \ac{HT} provides two ``logical''
processors for each physical core. Studies from Intel and others have
shown that~\ac{HT} contributes a performance boost between
$16$--$28\%$~\cite{HT}. This means that when two processes are
executed on the same core, the processes compete for the shared
hardware resources resulting is lower efficiency. 

When running an OpenMP program it is possible to choose the number $P$
of threads to use, either in the source code or through the
\verb+OMP_NUM_THREADS+ environment variable. In our experiments below,
$P$ never exceeds twice the number of physical cores provided by the
processor, so that the \ac{OS} will be able to assign each thread to a
separate (logical) core. Unless configured differently, the Linux
scheduler tries to spread processes to different cores as far as
possible; only when there are more runnable processes than cores does
\ac{HT} come into effect.

For better comparability of our results with those reported in the
literature we consider $d=1$ dimensions and use the methodology and
parameters described in~\cite{Raczy2005}. The first parameter is the
total number of extents~$N$, that includes $n=N/2$ subscription and
$m=N/2$ update extents. All extents are randomly placed on a segment
of total length $L=10^6$ and have the same length $l$. The segment
length is defined in such a way that a given overlapping degree $\alpha$
is obtained, where

\begin{equation*}
\alpha=\frac{\sum \text{area of extents}}{\text{area of the routing space}} = \frac{N \times l}{L}
\end{equation*}

Therefore, given $\alpha$ and $N$, the length $l$ of each segment is
set to $l = \alpha L / N$. The overlapping degree is an indirect
measure of the total number of intersections among subscription and
update extents. While the cost of~\ac{BF} and~\ac{SBM} is not affected
by the number of intersections, this is not the case for~\ac{ITM}, as
will be shown below. We considered the same values for $\alpha$ as
in~\cite{Raczy2005}, namely $\alpha \in \{0.01, 1, 100\}$.  Finally,
each measure is the average of~$30$ independent runs to get
statistically valid results.  Our implementations do not explicitly
store the list of intersections, but only count them. We did so to
ensure that the algorithms run time is not affected by the choice of
the data structure used to store the intersections.

\begin{figure*}[ht]
  \centering
  \subfigure[\label{fig:wct-n}]{\includegraphics[width=.95\columnwidth]{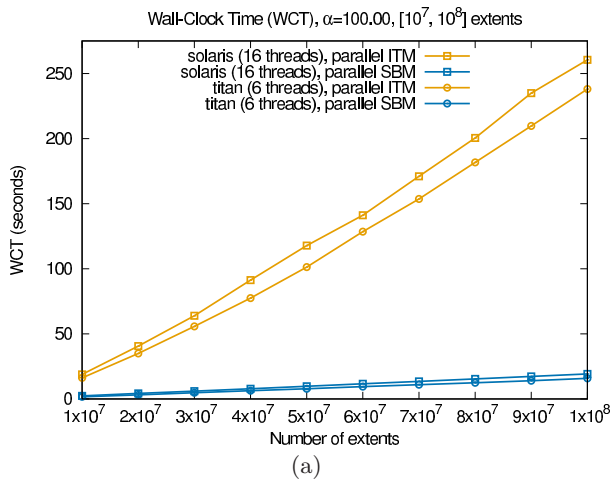}}\quad
  \subfigure[\label{fig:wct-alpha}]{\includegraphics[width=.95\columnwidth]{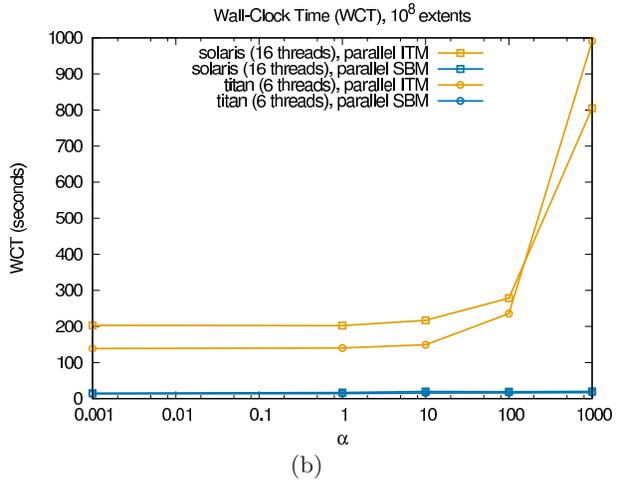}}
  \caption{Wall clock time of parallel~\ac{ITM} and \ac{SBM} as a
    function of the number of extents $N$~(a) and overlapping degree
    $\alpha$~(b) (note the logarithmic horizontal
    scale).}\label{fig:wct-n-alpha}
\end{figure*}

\paragraph*{Wall clock time}
The first performance metric we analyze is the \ac{WCT} of the
algorithms. Figure~\ref{fig:wct-1M} shows the \ac{WCT} for the
parallel versions of \ac{BF}, \ac{ITM} and \ac{SBM} as a function of
the number $P$ of OpenMP threads used, given $N=10^6$ extents and
overlapping degree $\alpha = 100$. Dashed lines indicate when $P$
exceeds the number of CPU cores.

With those parameters, the parallel~\ac{BF} algorithm is about three
orders of magnitude slower than \ac{SBM} on both the \texttt{titan}
and \texttt{solaris} machine. For larger values of $N$ the gap widens
further, since \ac{BF} is asymptotically slower than the other two
algorithms. Indeed, the computational cost of \ac{BF} grows
quadratically with the number of extents (see
Section~\ref{sec:region-matching}), while that of \ac{SBM} and
\ac{ITM} grows only polylogarithmically. \ac{ITM} performs better than
\ac{BF}, but worse than \ac{SBM}.

In Figure~\ref{fig:wct-n-alpha} we study how the~\ac{WCT} of the
parallel \ac{ITM} and \ac{SBM} algorithms depend on the number of
extents $N$ and the overlapping degree $\alpha$. The measures were
taken on both machines (\texttt{solaris} and \texttt{titan}) with as
many OpenMP threads as physical cores. Figure~\ref{fig:wct-n} shows
that the \ac{WCT} grows polylogarithmically with $N$ for both \ac{ITM}
and \ac{SBM}, confirming the asymptotic analysis in
Section~\ref{sec:parallel-sort-matching}; however, the parallel
\ac{SBM} algorithm is faster than \ac{ITM} on both machines,
suggesting that its asymptotic cost has smaller constants and terms of
lower order. 

In Figure~\ref{fig:wct-alpha} we report the \ac{WCT} as a function of
$\alpha$, for a fixed $N=10^8$. We observe that, unlike \ac{ITM}, the
execution time of \ac{SBM} is essentially independent from the
overlapping degree.

\paragraph*{Speedup}
The \emph{relative speedup} measures the increase in speed that a
parallel program achieves when more processors are employed to solve a
problem of the same size. This metric can be computed from the
\ac{WCT} as follows. Let $T(N, P)$ be the \ac{WCT} required to process
an input of size $N$ using $P$ processes (OpenMP threads). Then, for a
given $N$, the relative speedup $S_N(P)$ is defined as $S_N(P) = T(N,
1) / T(N, P)$. Ideally, the maximum value of $S_N(P)$ is $P$, which
means that solving a problem with $P$ processors requires $1/P$ the
time needed by a single processor. In practice, however, several
factors limit the speedup, such as the presence of serial regions in
the parallel program, uneven load distribution, scheduling overhead,
and heterogeneity in the execution hardware.

Figure~\ref{fig:speedup-1M} shows the speedups of the parallel
versions of \ac{BF}, \ac{ITM} and \ac{SBM} as a function of the number
of OpenMP threads $P$; the speedups have been computed using the wall
clock times of Figure~\ref{fig:wct-1M}. Line colors denote the
algorithm, while the shape of the data points denote the host where
the tests have been executed (square = \texttt{solaris}, circle =
\texttt{titan}). Dashed lines indicate data points where $P$ exceeds
the number of physical processor cores available on that machine.

The~\ac{BF} algorithm, despite being the less efficient, is the most
scalable. This can be attributed to its embarrassingly parallel
structure and lack of any serial part. \ac{SBM}, on the other hand, is
the most efficient but the less scalable. Interestingly, with equal
number of OpenMP threads, \ac{SBM} and \ac{ITM} scale better on the i7
machine (\texttt{titan}) than on the Xeon machine (\texttt{solaris}),
while \ac{BF} seems unaffected by the processor type.  \ac{SBM}
achieves a $2.6\times$ speedup with $16$ OpenMP threads on the dual
Xeon machine, and a $2.9\times$ speedup with $6$ OpenMP threads on the
Core i7 machine.  When all ``virtual'' cores are used, the speedup
grows to $3.6\times$ on the Xeon machine and $4.1\times$ on the i7.

The effect of \ac{HT} (dashed lines) is clearly visible in
Figure~\ref{fig:speedup-1M}. The speedup degrades when $P$ exceeds the
number of cores, as can be seen from the different slopes for \ac{BF}
on \texttt{titan}. When \ac{HT} kicks in, load unbalance arises due to
contention of the shared control units of the processor cores, and
this limits the scalability. The bizarre behavior of \ac{BF} on
\texttt{solaris} around $P=24$ (the speedup drops and then raises
again) is likely caused by OpenMP \ac{NUMA} scheduling
issues~\cite{Broquedis2008}. Considering the high wall clock time of
the \ac{BF} algorithm, we do not address this issue in this paper.

\begin{figure}[ht]
  \centering%
  \includegraphics[width=.95\columnwidth]{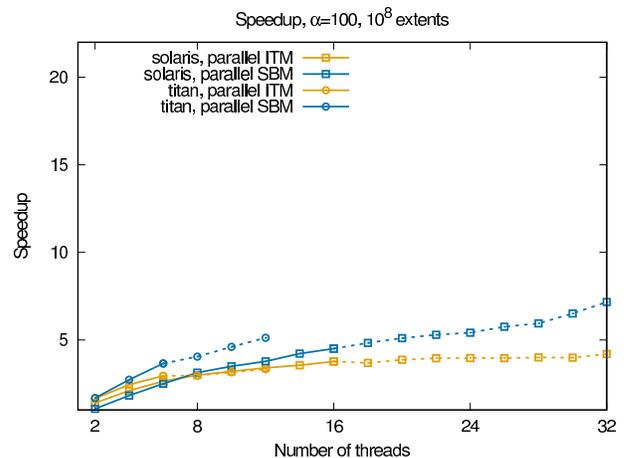}
  \caption{Speedup of parallel~\{\ac{ITM},~\ac{SBM}\} with $N=10^8$
    extents, overlapping degree $\alpha=100$.}\label{fig:speedup-100M}
\end{figure}

The speedup of~\ac{SBM} improves slightly if we increase the work
performed by the algorithm. Figure~\ref{fig:speedup-100M} shows the
speedup of parallel \ac{ITM} and \ac{SBM} with $N=10^8$ extents and
overlapping degree $\alpha = 100$ (in this scenario \ac{BF} takes so
long that it has been omitted). The~\ac{SBM} algorithm behaves better,
especially on the dual socket Xeon machine, achieving a $4.5\times$
speedup with $16$ OpenMP threads, and $7\times$ speedup with $32$
threads.  On the Core i7 machine the speedup is $3.6\times$ with $6$
OpenMP threads (one per core), and $5.1\times$ with $12$ threads (two
per core).

\begin{figure*}[t]
  \centering%
  \subfigure[Strong scaling\label{fig:strong-scaling}]{\includegraphics[width=.95\columnwidth]{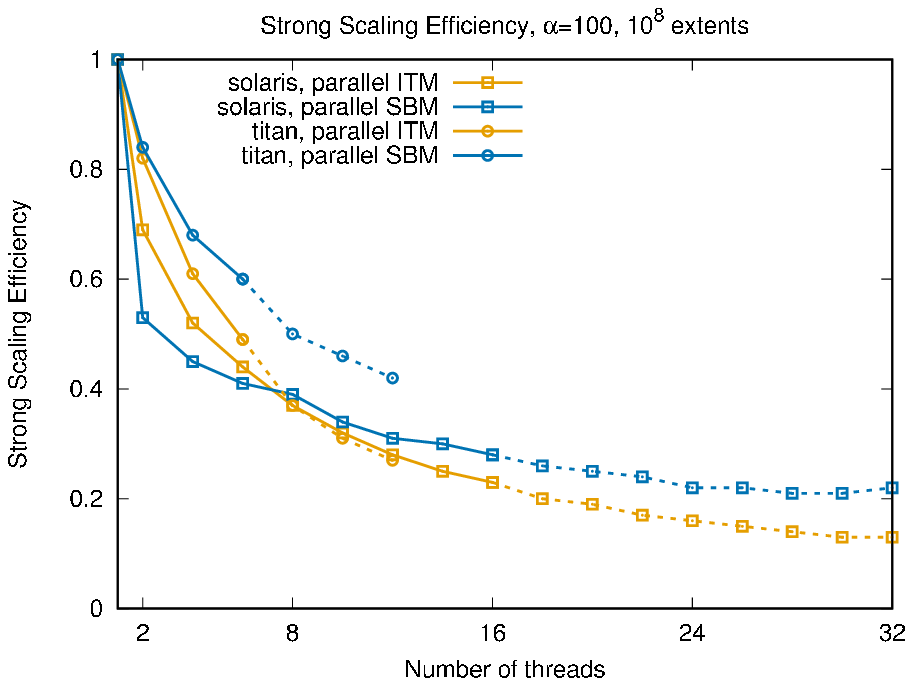}}\quad
  \subfigure[Weak scaling\label{fig:weak-scaling}]{\includegraphics[width=.95\columnwidth]{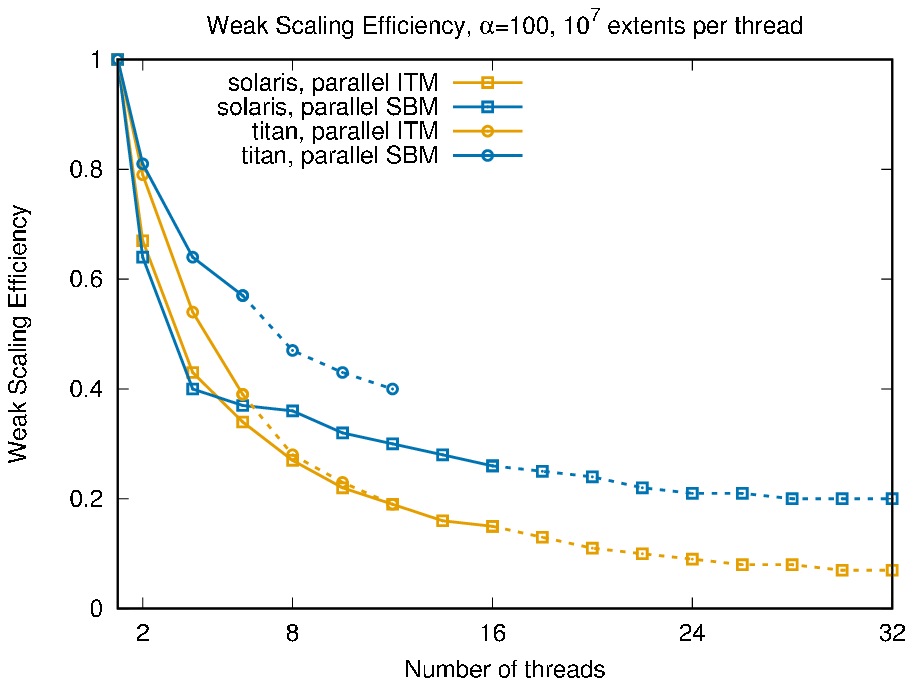}}
  \caption{Strong and weak scaling behavior of parallel~\ac{ITM}
    and~\ac{SBM}, overlapping degree $\alpha=100$.}\label{fig:scaling}
\end{figure*}

\paragraph*{Scaling Efficiency}
The scaling efficiency measures how well a parallel application
exploits the available computational resources. Two formulation of
scaling efficiency are given in the literature: \emph{strong scaling}
and \emph{weak scaling}. Given an input of size $N$ and $P$
processors, the strong scaling efficiency $E_{N, \textrm{strong}}(P)$
and weak scaling efficiency $E_{N, \textrm{weak}}(P)$ are defined as:
\begin{align*}
E_{N, \textrm{strong}}(P) &= \frac{T(N, 1)}{P \times T(N, P)} = \frac{S_N(P)}{P} \\
E_{N, \textrm{weak}}(P) &= \frac{T(N, 1)}{T(P \times N, P)}
\end{align*}
Scaling efficiencies are real numbers in the range $[0, 1]$. An
efficiency of, say, $0.8$ indicates that the application spends $80\%$
of the time doing actual work, the rest being communication and
synchronization overhead. Therefore, higher efficiencies denote better
scaling behavior. Strong scaling measures how well a parallel
application exploits the processors, assuming constant total problem
size. Weak scaling measures how well the application exploits the
processors under constant per-processor work.

Strong and weak scaling are investigated in Figure~\ref{fig:scaling},
assuming overlapping degree $\alpha = 100$.  We observe that both
efficiencies sharply drop when going from $P=1$ to $P=2$ and $P=4$
OpenMP threads. Looking at the strong scaling behavior of \ac{SBM} and
\ac{ITM} (Figure~\ref{fig:strong-scaling}), \ac{SBM} scales better
than \ac{ITM} on the Intel i7 machine. On the other hand, \ac{ITM} is
more efficient than \ac{SBM} on the Xeon machine up to $P=8$ OpenMP
threads; after that, \ac{SBM} becomes more efficient. Weak scaling
(Figure~\ref{fig:weak-scaling}) shows a similar behavior: \ac{SBM}
scales consistently better than \ac{ITM} on \texttt{titan}, while on
\texttt{solaris} \ac{ITM} is better than \ac{SBM} up to $P=6$ OpenMP
threads. 

Figure~\ref{fig:scaling} confirms that the OpenMP \ac{SBM}
implementation exhibits efficiency issues. The reason is still being
investigated, since it is unclear whether \ac{NUMA} memory issues can
explain this behavior.

\begin{figure*}[t]
  \centering%
  \subfigure[\label{fig:memory-extents}]{\includegraphics[width=.95\columnwidth]{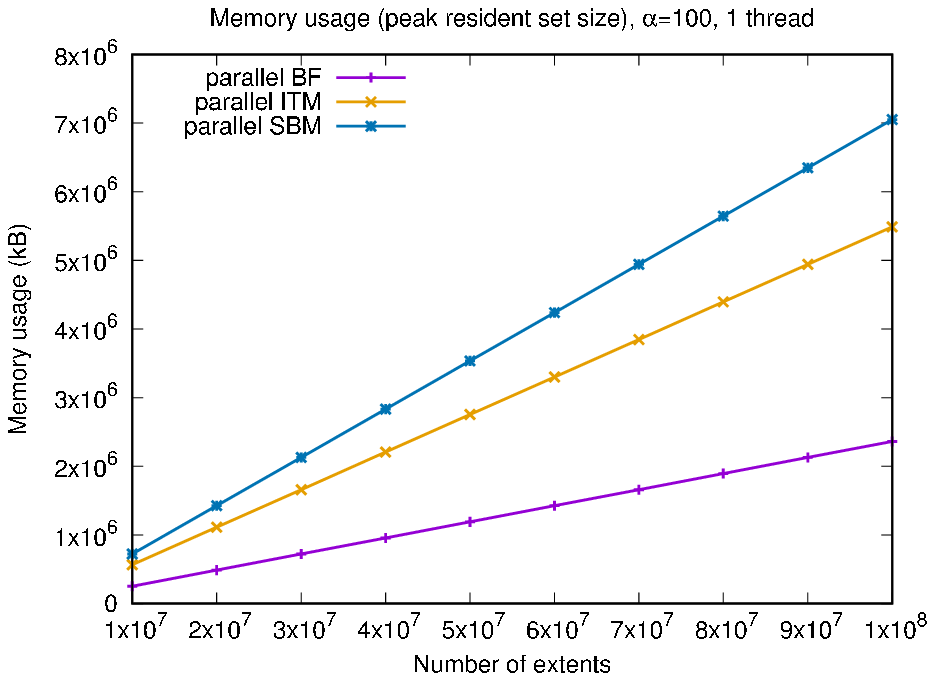}}\quad
  \subfigure[\label{fig:memory-threads}]{\includegraphics[width=.95\columnwidth]{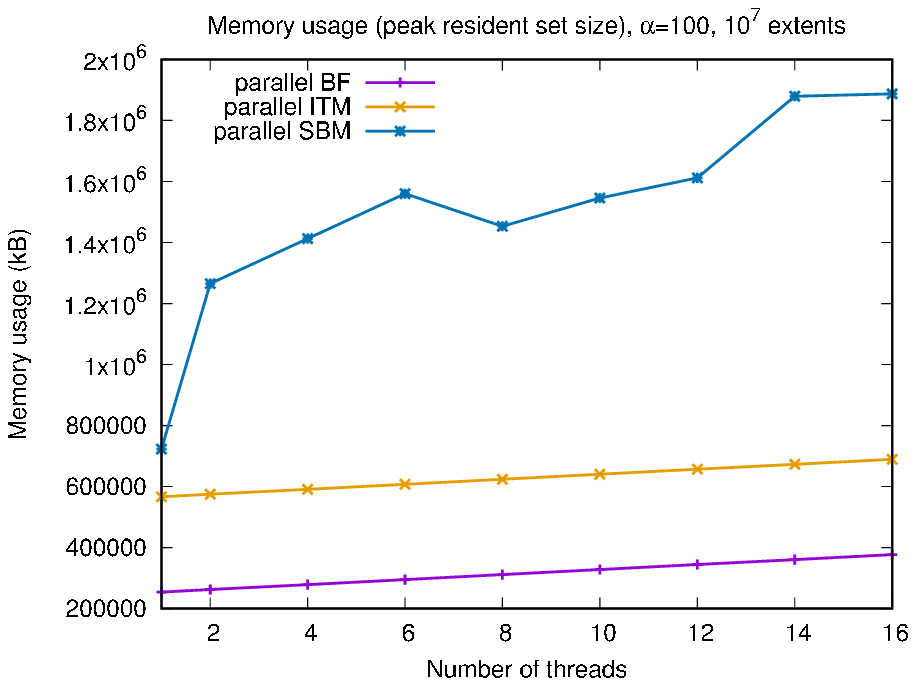}}
  \caption{Memory usage (peak resident set size, VmHWM) of parallel~\{\ac{BF},~\ac{ITM},~\ac{SBM}\} with an increasing number of extents~(a) or threads~(b),
    overlapping degree $\alpha=100$.}\label{fig:memory}
\end{figure*}

\paragraph*{Memory Usage}
We conclude our experimental evaluation with an assessment of the
memory usage of the parallel \ac{BF}, \ac{ITM} and \ac{SBM}
algorithms. Figure~\ref{fig:memory} shows the peak \ac{RSS} of the
three algorithms as a function of the number of extents and OpenMP
threads; the data have been collected on the Xeon machine
\texttt{solaris}. The \ac{RSS} is the portion of a process memory that
is kept in RAM. Care has been taken to ensure that all experiments
reported in this section fit comfortably in the main memory of the
available machines, so that the \ac{RSS} represents an actual upper
bound of the amount of memory required by the algorithms. Note that
the data reported in Figure~\ref{fig:memory} includes the code for the
test driver and the input arrays of intervals.

Figure~\ref{fig:memory-extents} shows that the resident set size grows
linearly with the number of extents $N$. \ac{BF} has the smaller
memory footprint, since it requires a tiny bounded amount of
additional memory for a few local variables; \ac{SBM} uses more memory
since it allocates larger data structures, namely the list of
endpoints to be sorted, and a few arrays of sets. \ac{SBM} requires
approximately $7$~GB of memory to process $N=10^8$ intervals, about
three times the amount of memory required by \ac{BF}.

In Figure~\ref{fig:memory-threads} we study the \ac{RSS} as a function
of the number of OpenMP threads $P$. The \ac{RSS} for \ac{BF} and
\ac{ITM} grows very slowly with $P$, since they do not explicitly use
additional per-thread variables; therefore, the marginal increase of
\ac{RSS} that we observe is due to the normal overhead of the OpenMP
threading system. On the other hand, the \ac{RSS} for the parallel
\ac{SBM} algorithm is strongly influenced by the number of threads,
although the variability is so high that it is not possible to observe
a smooth correlation (despite each data point being computed over
multiple runs). Such variability is likely caused by memory
fragmentation induced by the \ac{STL} data structures used by the
algorithm. In any case, the \ac{RSS} for \ac{SBM} shows only a
threefold increase when moving fro $P=2$ to $P=16$ OpenMP threads; we
can therefore postulate that the \ac{RSS} will not become a bottleneck
for any reasonable number of OpenMP threads that are used.

\section{Conclusions and Future Works}\label{sec:conclusions}

In this paper we described a parallel version of the \ac{SBM}
algorithms for solving the $d$-rectangle intersection problem for
\acl{DDM}. Our algorithm is targeted at shared-memory multicore
architectures that constitute the vast majority of current processors.

We have implemented the parallel \ac{SBM} algorithm in C++ with OpenMP
directives. Performance measurement shows a $2.6\times$ speedup
with~$16$ OpenMP threads on a dual Intel Xeon processor, and a
$2.9\times$ speedup with~$6$ threads on an Intel Core i7
processor. The memory usage of parallel \ac{SBM} grows linearly with
the number of extents $N$; memory usage also depends on the number of
OpenMP threads used. In any case, \ac{SBM} uses about $7$~GB of memory
to handle~$100$ millions of extents, making it attractive for large
scenarios.

We are currently extending the present work in two directions. First,
we are studying how the choice of parallel sorting algorithm and of
dynamic set data structure influence the scalability of parallel
\ac{SBM}. Indeed, while the results presented in
Section~\ref{sec:experimental-evaluation} are encouraging, scalability
is lower than what asymptotic analysis predicts, suggesting the
presence of bottlenecks in the implementation that should be
identified and removed. Scheduling issues and \ac{NUMA} memory
conflicts are suspected to play a significant role in the loss of
efficiency that we observe on the dual-socket test machine. Second, we
are extending the \ac{SBM} algorithm to solve the dynamic~\ac{DDM}
matching problem, where extents can be moved or resized
dynamically. This problem has already been investigated in the context
of serial \ac{SBM}~\cite{Pan2011}, so it is important to assess if and
how it can be solved in a parallel environment.

\section*{Notation}

\begin{center}
\begin{tabular}{rl}
$\mathbf{S}$ & Subscription set $\mathbf{S} = \{S_1, \ldots, S_n\}$ \\
$\mathbf{U}$ & Update set $\mathbf{U} = \{U_1, \ldots, U_m\}$ \\
$n$ & Number of subscription extents \\
$m$ & Number of update extents\\
$N$ & N. of subscription and update extents ($N = n + m$)\\
$K$ & Number of intersections, $0 \leq K \leq n \times m$ \\
$\alpha$ & Overlapping degree ($\alpha > 0$)\\
$P$ & Number of processors/OpenMP threads\\
\end{tabular}
\end{center}

\section*{Acronyms}

\begin{acronym}[NUMA]
\acro{BF}{Brute Force}
\acro{DDM}{Data Distribution Management}
\acro{GB}{Grid Based}
\acro{HLA}{High Level Architecture}
\acro{HT}{Hyper-Threading}
\acro{ITM}{Interval Tree Matching}
\acro{NUMA}{Non Uniform Memory Access}
\acro{OS}{Operating System}
\acro{RSS}{Resident Set Size}
\acro{SBM}{Sort-based Matching}
\acro{STL}{Standard Template Library}
\acro{WCT}{Wall Clock Time}
\end{acronym}

\bibliographystyle{abbrv}
\bibliography{parallel-sort-based-matching.bib}

\begin{thebibliography}{10}

\bibitem{HLA}
{IEEE Standard for Modeling and Simulation (M\&S) High Level Architecture
  (HLA)--Framework and Rules}.
\newblock {IEEE Std 1516-2010 (Rev. of IEEE Std 1516-2000)}, 2010.

\bibitem{avl}
G.~Adelson-Velskii and E.~M. Landis.
\newblock {An Algorithm for the Organization of Information}.
\newblock {\em Doklady Akademii Nauk USSR}, 146(2):263--266, 1962.

\bibitem{Bayer1972}
R.~Bayer.
\newblock Symmetric binary {B-Trees}: Data structure and maintenance
  algorithms.
\newblock {\em Acta Informatica}, 1(4):290--306, 1972.

\bibitem{Blelloch89}
G.~E. Blelloch.
\newblock Scans as primitive parallel operations.
\newblock {\em IEEE Transactions on Computers}, 38(11):1526--1538, Nov 1989.

\bibitem{Boukerche2001}
A.~Boukerche and C.~Dzermajko.
\newblock Performance comparison of data distribution management strategies.
\newblock In {\em Proc. 5th IEEE Int. Workshop on Distributed Simulation and
  Real-Time Applications}, DS-RT '01, pages 67--, Washington, DC, USA, 2001.
  IEEE Computer Society.

\bibitem{Boukerche2002}
A.~Boukerche and A.~Roy.
\newblock Dynamic grid-based approach to data distribution management.
\newblock {\em Journal of Parallel and Distributed Computing}, 62(3):366--392,
  2002.

\bibitem{Broquedis2008}
F.~Broquedis, F.~Diakhat{\'e}, S.~Thibault, O.~Aumage, R.~Namyst, and P.-A.
  Wacrenier.
\newblock Scheduling dynamic openmp applications over multicore architectures.
\newblock In R.~Eigenmann and B.~R. de~Supinski, editors, {\em OpenMP in a New
  Era of Parallelism: 4th International Workshop, IWOMP 2008 West Lafayette,
  IN, USA, May 12-14, 2008 Proceedings}, pages 170--180, Berlin, Heidelberg,
  2008. Springer Berlin Heidelberg.

\bibitem{parallel-stl}
Programming languages -- technical specification for {C++} extensions for
  parallelism.
\newblock ISO/IEC TS 19570:2015, 2015.

\bibitem{Cole88}
R.~Cole.
\newblock Parallel merge sort.
\newblock {\em SIAM Journal on Computing}, 17(4):770--785, 1988.

\bibitem{OpenMP}
L.~Dagum and R.~Menon.
\newblock {OpenMP}: An industry-standard {API} for shared-memory programming.
\newblock {\em IEEE Comput. Sci. Eng.}, 5:46--55, January 1998.

\bibitem{gda-simpat}
G.~D'Angelo and M.~Marzolla.
\newblock New trends in parallel and distributed simulation: From many-cores to
  cloud computing.
\newblock {\em Simulation Modelling Practice and Theory}, 49:320--335, 2014.

\bibitem{Devai2010}
F.~Devai and L.~Neumann.
\newblock A rectangle-intersection algorithm with limited resource
  requirements.
\newblock In {\em Proc. 10th IEEE Int. Conf. on Computer and Information
  Technology}, CIT '10, pages 2335--2340, Washington, DC, USA, 2010. IEEE
  Computer Society.

\bibitem{Guttman1984}
A.~Guttman.
\newblock R-trees: a dynamic index structure for spatial searching.
\newblock {\em SIGMOD Rec.}, 14(2):47--57, June 1984.

\bibitem{Jun2002}
Y.~Jun, C.~Raczy, and G.~Tan.
\newblock Evaluation of a sort-based matching algorithm for {DDM}.
\newblock In {\em Proceedings 16th Workshop on Parallel and Distributed
  Simulation}, pages 62--69, 2002.

\bibitem{liu2013}
E.~S. Liu and G.~K. Theodoropoulos.
\newblock An analysis of parallel interest matching algorithms in distributed
  virtual environments.
\newblock In {\em 2013 Winter Simulations Conference (WSC)}, pages 2889--2901,
  Dec 2013.

\bibitem{Yanbing2015}
Y.~Liu, H.~Sun, W.~Fan, and T.~Xiao.
\newblock A parallel matching algorithm based on order relation for hla data
  distribution management.
\newblock {\em International Journal of Modeling, Simulation, and Scientific
  Computing}, 06(02):1540002, 2015.

\bibitem{HT}
D.~T. Marr, F.~Binns, D.~L. Hill, G.~Hinton, D.~A. Koufaty, A.~J. Miller, and
  M.~Upton.
\newblock {Hyper-Threading Technology Architecture and Microarchitecture}.
\newblock {\em Intel Technology Journal}, 6(1), Feb. 2002.

\bibitem{gda-dsrt-2013}
M.~Marzolla, G.~D'Angelo, and M.~Mandrioli.
\newblock A parallel data distribution management algorithm.
\newblock In {\em 2013 IEEE/ACM 17th International Symposium on Distributed
  Simulation and Real Time Applications (DS-RT)}, pages 145--152, Oct 2013.

\bibitem{Medidi98}
M.~Medidi and N.~Deo.
\newblock {Parallel Dictionaries Using AVL Trees}.
\newblock {\em Journal of Parallel and Distributed Computing}, 49(1):146--155,
  1998.

\bibitem{Pan2011}
K.~Pan, S.~J. Turner, W.~Cai, and Z.~Li.
\newblock A dynamic sort-based {DDM} matching algorithm for {HLA} applications.
\newblock {\em ACM Trans. Model. Comput. Simul.}, 21(3):17:1--17:17, Feb. 2011.

\bibitem{Park01}
H.~Park and K.~Park.
\newblock Parallel algorithms for red-black trees.
\newblock {\em Theor. Computer Science}, 262(1):415--435, 2001.

\bibitem{petty-1997}
M.~D. Petty and A.~Mukherjee.
\newblock Experimental comparison of $d$-rectangle intersection algorithms
  applied to {HLA} data distribution.
\newblock In {\em Proceedings of the 1997 Distributed Simulation Symposium},
  pages 13--26, 1997.

\bibitem{Raczy2005}
C.~Raczy, G.~Tan, and J.~Yu.
\newblock A sort-based {DDM} matching algorithm for {HLA}.
\newblock {\em ACM Trans. Model. Comput. Simul.}, 15(1):14--38, Jan. 2005.

\bibitem{rao2013}
D.~Rao, X.~Hu, and L.~Wu.
\newblock Performance analysis of parallel data distribution management in
  large-scale battlefield simulation.
\newblock In {\em Proceedings of 2013 3rd International Conference on Computer
  Science and Network Technology}, pages 134--138, Oct 2013.

\bibitem{Rosenberg1985}
J.~Rosenberg.
\newblock Geographical data structures compared: A study of data structures
  supporting region queries.
\newblock {\em Computer-Aided Design of Integrated Circuits and Systems, IEEE
  Transactions on}, 4(1):53--67, 1985.

\bibitem{Stroustrup13}
B.~Stroustrup.
\newblock {\em {The C++ Programming Language, 4th Edition}}.
\newblock Addison-Wesley, 2013.

\bibitem{Wheat92}
M.~Wheat and D.~Evans.
\newblock An efficient parallel sorting algorithm for shared memory
  multiprocessors.
\newblock {\em Parallel Computing}, 18(1):91--102, 1992.

\end{thebibliography}

\end{document}